**RESEARCH ARTICLE**

# Extended Weighted ABG: A Robust Non-Linear ABG-Based Approach for Optimal Combination of ABG Path-Loss Propagation Models

**DAVID CASILLAS-PÉREZ**[1], **DANIEL MERINO-PÉREZ**[2], **SILVIA JIMÉNEZ-FERNÁNDEZ**[2], **JOSE A. PORTILLA-FIGUERAS**[2], AND **SANCHO SALCEDO-SANZ**[2]
[1]Department of Signal Processing and Communications, Universidad Rey Juan Carlos, Fuenlabrada, 28942 Madrid, Spain
[2]Department of Signal Processing and Communications, Universidad de Alcalá, 28801 Madrid, Spain

Corresponding author: David Casillas-Pérez (david.casillas@urjc.es)

This work was supported by the Universidad de Alcalá-ISDEFE Chair of Research in ICT and Digital Progress.

**ABSTRACT** This paper proposes a robust non-linear generalized path-loss propagation model, the Extended Weighted ABG (EWABG), which efficiently allows generating a path-loss propagation model by combining several available path-loss datasets (from measurements campaigns) and other previously proposed state-of-the-art 5G path-loss propagation models. The EWABG model works by integrating individual path-loss models into one single model in the least-squares sense, allowing to extend knowledge from frequencies and distances covered by path-loss datasets or path-loss propagation models. The proposed EWABG model is the first non-linear extension of the common ABG-based approach, which surpasses the non-uniformity problem between the low and high 5G frequencies (as most measurements campaigns have taken place in low frequencies). The EWABG also addresses the problem of removing outlier measurements, a step not included in previous propagation path-loss models. In this case, we have compared the most recent techniques for avoiding outliers, and we have adopted the Theil-Sen method, due to its strong robustness demonstrated in the experiments carried out. In addition, the proposed model specifically considers non-linear attenuation by atmospheric gases, in order to improve its estimations. The good performance of the proposed EWABG model has been tested and compared against recent 5G propagation path-loss models including the ABG and WABG models. The exhaustive experimentation carried out includes the 5G non-line-of-sight environment in different 5G scenarios, UMiSC, UMiOS and UMa. The proposed EWABG obtains the best accuracy, specially in noisy environments with outliers, reporting negligible increment error rates (with respect to the non-outliers situation), lower than 1%, compared to the ABG and WABG.

## I. INTRODUCTION

The fifth generation of mobile communications systems, more commonly known as 5G, is revolutionizing a multitude of sectors, such as business [1], [2], smart cities [3], IoT [4], health [5] or industry [6] and experiences [7]. This new technology is bringing about a complete transformation of modern communication systems, thanks to it characteristics such as 10 Gbps downlink throughput, 1 ms latency, 99.999% reliability and a large number of devices connected simultaneously to the access network [8], [9].

Providing these demanded high data rates and quality parameters, requires innovative technologies, such as non-orthogonal multiple access (NOMA) [10]–[14] and massive multiple-input multiple-output (mMIMO) [15]–[18], not previously included in the fourth generation of mobile communications (4G). Despite the considerable increase in data rates and quality provided by these techniques, the frequency bands allocated for mobile communication systems will not suffice for the needs of the new 5G technologies. In fact, the Third Generation Partnership Project (3GPP [9]),

The associate editor coordinating the review of this manuscript and approving it for publication was Muhammad Zubair.

suggests the need to use mmWave frequency bands for hosting future 5G services [19], [20].

One of the most important issues of using mmWave frequencies for mobile communications is the high propagation path-losses that they present [21]–[25]. Radio propagation modelling is one of the key points for designing mobile communication systems. Propagation models estimate the path-losses which are used to calculate the maximum coverage range of radio base stations, and the number of cells to completely cover a given area. In the literature, there exist three main types of radio propagation models [26]: ray tracing or optical models [27]–[29], dominant path models [30] and statistical models [25], [31], [32]. Ray tracing-optical models estimate path-losses by integrating the individual contributions of multiple rays between the transmitter and the receiver. These models are usually very accurate, since they solve the scattering, reflection and diffraction laws for each individual ray, with the consequent computational cost. Due to this high computation burden, these models often require to specify lots of information about the area under study, including geographic information (position of streets and buildings), as well as building construction materials. Thus, propagation models based on ray tracing have an extremely high computational cost, need large simulators and complete information about the structure of buildings and surroundings. Therefore, they are not fit to be implemented on base stations.

On the other hand, Dominant path models considerably reduce the computational cost, by considering the main contributing ray only. They also require specific information on the area and the area's materials [33], but only that concerning to the main contributing ray.

Finally, statistical models, also known as empirical models, provide accurate estimations of the propagation path-losses by using several environment parameters. The statistical models generally solve an optimization problem that involves different criteria, such as, maximum likelihood or least-squares. As a result, the models find the parameters that best fit the path-loss observations. These models are very useful when the available area information is reduced or when the computation time requirements are limited, as in the maximum cell range problem to be solved in mobile communications.

In the literature, there exist multiple statistical propagation path-loss models for 5G systems systems [9], [22], [31], [34]–[38]. In recent years, Machine Learning (ML) techniques have been used to model the 5G channel [39], [40], especially in indoor environments [41]. ML methods are included in the statistical category, because their parameters are usually estimated from samples (datasets). They are more accurate in general, but require large databases to be trained. In addition they require more computation resources than conventional statistical methods, and therefore they are in many cases unpractical for being used in the 5G radio layer in real-time. The most used path-loss model in 5G applications is the Alpha-Beta-Gamma (ABG) model [9], [22], [31]. The ABG is a large-scale propagation path-loss model established by the 3GPP as the standard model for 5G [9]. Currently, we count on specific ABG models for different 5G scenarios, such as, Urban Micro (UMi) and Macro cells (UMa) and in different environments such as Line-of-Sight and Non-Line-of-Sight (LOS, and NLOS respectively). There are different ABG models that cover a wide distance-frequency range, many of which overlap in terms of distance or frequency [22], [31]. To name some examples, [42] studied the wideband channels at 9.6, 28.8 and 57.6 GHz in LOS and NLOS. In [43] and [44] a measurement campaign and path-loss modeling was carried out in the 60 GHz band, and in [45] Aalto University studied frequencies in the E-band, specifically 71 to 76 GHz and 81 to 86 GHz bands. Although all of these models estimate similar path-loss values, especially in LOS (where propagation path-loss modelling is close to the free-space modeling), some dissimilarities may occur, and policies for balancing the path-loss approaches are needed to create an integrated model.

There are few integrated approaches that deal with the existent ABG models, and cover the 5G distance-frequency ranges. The first works [31] try to create a new integrated path-loss model by collecting different 5G path-loss databases into a new one. Then, this big database is used to compute a new set of ABG coefficients. The procedure has some troubleshoots: first, data are collected into a new dataset without a balancing or weighting policy. As a consequence, the integrated ABG model tends to overestimate those models with a larger number of observations. Second, the procedure also requires the original path-losses datasets of each original ABG model in order to integrate them into a new one, which are difficult to find since they are not frequently public databases. Many works, see [31], only provide the resulting ABG coefficients of a model, their working ranges and a noise figure indicating how accurate the model is with respect to its database.

The Weighted ABG model (WABG) [25] was recently proposed to surpass the aforementioned drawbacks of the integrated ABG approaches. It is an ABG-based model which preserves the low number of parameters, but includes balanced policies based on different weighting criteria. Besides, the WABG model is able to combine different types of inputs: raw path-loss datasets and diverse ABG models provided in previous works. WABG helped to improve the accuracy of path-loss estimation, specially in NLOS environments.

Although, WABG overcomes the main limitations of the ABG model, it still has some drawbacks that prevent improving accuracy: First, ABG and WABG are both linear models. Considering the wide 5G working range and the differences across the domain, especially between high and low frequencies, a linear model will hardly accommodate them. Therefore, a non-linear model might fit better in these ranges, providing better path-loss estimations. The most obvious variations across the domain are caused by non-linearities due to the absorption of atmospheric gases during transmission, in particular oxygen and water. The absorption curves due to atmospheric gases have been known for decades [46], and their effect could be easily eliminated. Second, phenomena such as scattering, fading and noise, frequent in mobile communications, produce a large amount of outliers in data, which should be removed to improve the accuracy in the estimation of path-loss. Neither the ABG nor the WABG handle the problem of outlier presence in data, as both models are particularly sensitive to outliers, due to the fact that their coefficients are calculated through the L2-norm. Introducing methods that are statistically robust to outliers is essential to improve the path-loss models.

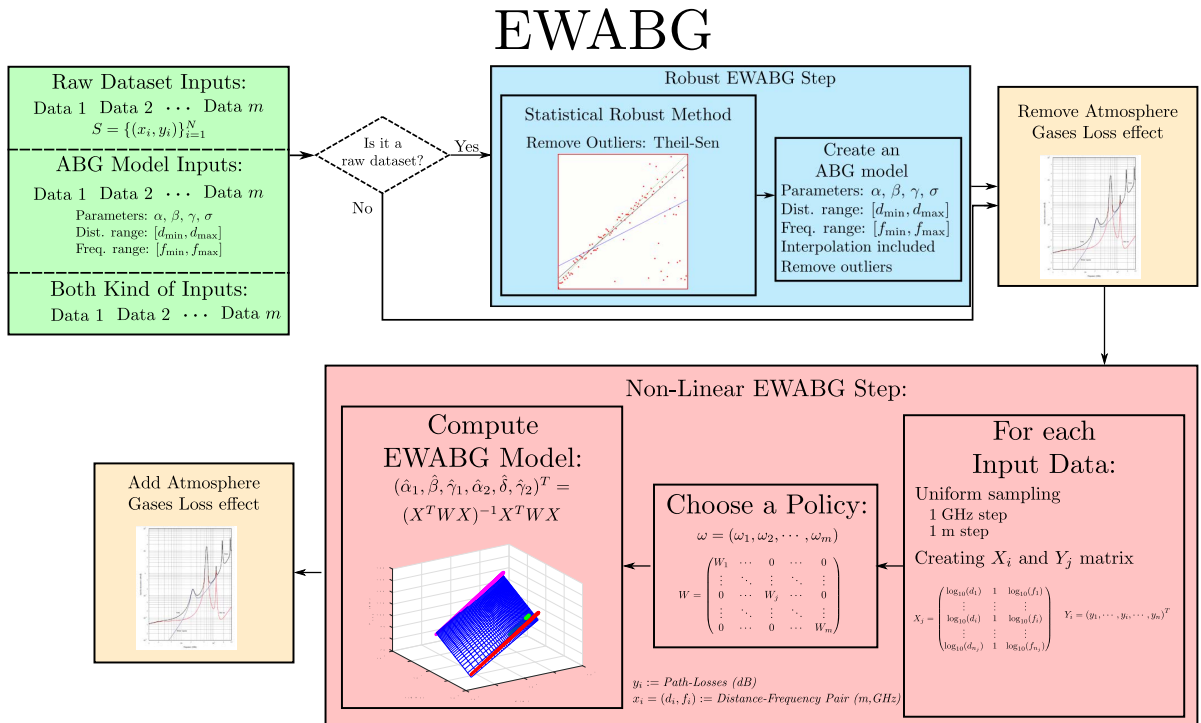

**FIGURE 1.** EWABG Flowchart. The proposed EWABG is a robust non-linear path-loss propagation model that integrates both raw datasets and other ABG models from different studies using a weighting policy.

With these previous discussion in mind, the main goal of the present work is to provide a non-linear propagation path-loss model, robust against outliers and able to combine different ABG models and/or raw path-loss datasets provided by different studies into one integrated ABG model. We call this new approach Extended WABG (EWABG), and its main characteristic is that it preserves the low number of parameters characteristic of the ABG models. To our knowledge, this work presents the first non-linear path-loss model which improves the path-loss estimations using few coefficients (only 6 coefficients), preserving this beneficial property from ABG-based model. Besides, the proposed model fashionably resolves the outliers presence problem, not addressed by any of the previous path-loss models in the literature. In EWABG, the integration of path-loss models or datasets into the integrated model follows the WABG weighting strategy, but with a non-linear procedure, that allows to surpass the aforementioned drawbacks of the state-of-the-art approaches. Figure 1 shows an illustrative workflow of the proposed EWABG model.

The proposed EWABG model has been assessed in several 5G scenarios, including UMiSC, UMiOS and UMa scenarios, in the critical NLOS environment, where the path-loss estimation is particularly difficult. The 5G LOS environment has been excluded from the experimental evaluation of the EWABG, since it does not offer significant room for improvement, see [25], [31]. We also compare the proposed EWABG model with several recently published ABG and WABG models in the considered 5G scenarios. We study environments with and without outliers, showing that the EWABG obtains excellent results in terms of reduced path-loss standard deviation.

The rest of the paper has been organized in different sections, as follows: Section II reviews the main characteristics of both ABG and WABG propagation path-loss models. Section III details the characteristics of the proposed EWABG model, including the non-linear terms added and the robust technique against outliers. Section IV first shows experiments for choosing the basis of the EWABG, specifically the order and the robust estimator, and finally it presents a complete set of experiments in NLOS environments for the different 5G considered scenarios. This section also shows a direct comparison with the state-of-the-art ABG models. Section V gives some final conclusions and future lines of research.

## II. THE ABG AND WABG MODELS

This section describes in detail the ABG and WABG path-loss models, their mathematical formulation and main drawbacks, which has led us to propose the EWABG as an improved path-loss model for 5G applications.

### A. THE ABG MODEL

One of the most extended 5G path-loss propagation model is the so called Alpha-Beta-Gamma model [9], or simply ABG model (named after the only three parameters it uses). It is a large-scale model, capable of representing the path-losses in a wide distance-frequency range. The ABG model is non-scenario and non-environment constraint, hence, it can represent path losses of both 5G LOS and NLOS environments in any 5G scenarios, such as UMi or UMa. The model expresses the path-losses as a linear combination of three coefficients $\alpha$, $\beta$ and $\gamma$, as follows:

$$P_{\text{ABG}}(d, f)[\text{dB}] = 10\alpha \log_{10}(d) + \beta + 10\gamma \log_{10}(f) + \xi \quad (1)$$

Here, the output $P_{\text{ABG}}$ is the path-loss estimation in decibels (dB). The formula shows a logarithmic dependence between the losses and both the transmitter-receiver distance $d$ and carrier frequency $f$ involved in the communication, expressed in meters (m) and Gigahertz (GHz), respectively. The parameters $\alpha$, $\beta$ and $\gamma$ are the coefficients of the ABG models that have to be estimated. The term $\xi$ is a random variable characteristic of the model which represents the path-losses fluctuations which appear along the transmission. It is assumed a zero mean and a finite variance $v^2$ for this fluctuation.

To obtain the $\alpha$, $\beta$ and $\gamma$ coefficients, the main existing approaches establish a linear least-squares problem to find the estimations that best fit a given dataset [25], [31], see Figure 2.

Let $X$ be the matrix that collects the distance-frequency observation points of the dataset, and let $Y$ be the vector that collects the corresponding path-loss measurements in these points, see Equation (2):

$$X = \begin{pmatrix} \log_{10}(d_1) & 1 & \log_{10}(f_1) \\ \vdots & \vdots & \vdots \\ \log_{10}(d_i) & 1 & \log_{10}(f_i) \\ \vdots & \vdots & \vdots \\ \log_{10}(d_n) & 1 & \log_{10}(f_n) \end{pmatrix}, \quad Y = \begin{pmatrix} y_1 \\ \vdots \\ y_i \\ \vdots \\ y_n \end{pmatrix}. \quad (2)$$

The optimal coefficients $(\hat{\alpha}, \hat{\beta}, \hat{\gamma})$ that minimize the sum of the squares of the residuals for linear models has a unique analytic formula, given by Equation (3):

$$\begin{pmatrix} \hat{\alpha} \\ \hat{\beta} \\ \hat{\gamma} \end{pmatrix} = (X^T X)^{-1} X^T Y. \quad (3)$$

Observe that the ABG model is able to model path-losses for a single carrier frequency by fixing its $\gamma$ coefficient to 0 or 2, as detailed in [32], [36], and [47]. The resulting model becomes a two-parameters model in $\alpha$ and $\beta$.

The ABG model frequently obtains accurate path-loss estimations when its coefficients are computed from one specific database. However, it struggles modelling path-losses from multiple databases which cover different distance-frequency ranges. The ABG model's accuracy lowers when input data come from multiple datasets and no weighting policy is applied if input datasets are unbalanced or there is data heteroscedasticity [25], [31].

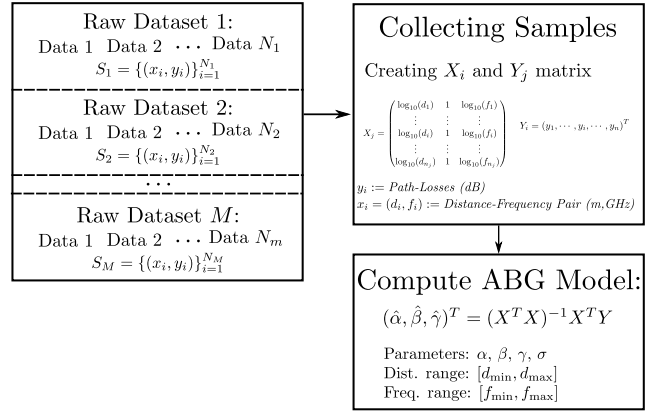

**FIGURE 2.** ABG flowchart.

### B. THE WABG MODEL

The Weighted ABG model [25], or simply WABG model, was proposed to overcome the previously described disadvantages of the ABG models when integrating multiple databases. WABG considers data heteroscedasticity and unbalance databases, in order to bind together multiple databases or path-loss models.

WABG distinguishes three different scenarios: Scenario a) considers several measurement campaigns as inputs. In other words, it assumes that the multiple databases to be integrated, which cover different distance-frequency ranges, are fully available. Scenario b) considers that no measurement campaign dataset is available, so it considers several input ABG models (determined by their $\alpha$, $\beta$ and $\gamma$ coefficients, the standard deviation and the distance-frequency working range). Scenario c) is a general case which combines the two previous scenarios, i.e., there are measurements for some distance-frequency ranges, and ABG coefficients for others.

WABG follows a similar procedure to ABG models to compute its coefficients [25]. Figure 3 illustrates its pipeline. It also establishes a linear least-squares problem, like a standard ABG approach, but weighting the residuals according to a given weighting policy. The best coefficient estimation $(\hat{\alpha}, \hat{\beta}, \hat{\gamma})$ has the following closed-form formula:

$$\begin{pmatrix} \hat{\alpha} \\ \hat{\beta} \\ \hat{\gamma} \end{pmatrix} = (X^T W X)^{-1} X^T W Y, \quad (4)$$

similar to expression (3), but involving a weighting matrix $W$. The weighting matrix $W$ is a block diagonal matrix whose diagonal components are composed by the corresponding

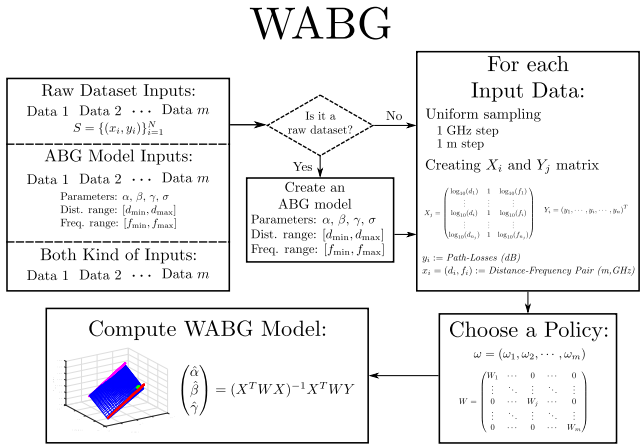

**FIGURE 3.** WABG flowchart.

weighting coefficients according to the applied policy [25]:

$$W = \begin{pmatrix} W_1 & \cdots & 0 & \cdots & 0 \\ \vdots & \ddots & \vdots & \ddots & \vdots \\ 0 & \cdots & W_j & \cdots & 0 \\ \vdots & \ddots & \vdots & \ddots & \vdots \\ 0 & \cdots & 0 & \cdots & W_m \end{pmatrix}, W_j = \begin{pmatrix} \frac{1}{\sigma_j^2} & \cdots & 0 & \cdots & 0 \\ \vdots & \ddots & \vdots & \ddots & \vdots \\ 0 & \cdots & \frac{1}{\sigma_j^2} & \cdots & 0 \\ \vdots & \ddots & \vdots & \ddots & \vdots \\ 0 & \cdots & 0 & \cdots & \frac{1}{\sigma_j^2} \end{pmatrix}. \quad (5)$$

WABG specifically proposes three main weighting policies: a) balancing the number of points in each dataset, b) weighting by the standard deviation of the fluctuation, and c) a mixture of both previous policies. Let us remark that the WABG model becomes a standard ABG model by substituting the weighting matrix by the identity matrix $W = I$. In this sense, the WABG is a generalization of the ABG path-loss model.

### C. ABG AND WABG DRAWBACKS

There are three main drawbacks that affect both ABG and WABG models when estimating path-losses.

1) Linear path-loss models: Both ABG and WABG models are linear and they estimate the path-losses by a linear combination that depends on the log-distance and the log-frequency, as can be seen in Equation (1), shared for both models. Of course, the non-homogeneous propagation medium prevents accurate path-losses estimations with these models, since it does not fit a plane in the logarithmic space of distance-frequency. Observations prove that there exist non-linear inconsistencies when their coefficients are computed in a wide working range (for a wide distance-frequency range). The most evident are the path-losses differences between high and low frequencies, specially in the NLOS environment, see [25]. Obtaining a more accurate model requires a non-linear path-loss model that takes into account this non-homogenous working range.
2) Sensitivity to outliers: None of the ABG or WABG models consider the possible existence of outliers during the data acquisition of the propagation path-loss databases. Outliers are those data points that significantly differ from other samples or observations. They occur due to the variability in the measurement campaign, given by the presence of interference, fading or scattering events during the transmission. Both models are specially sensitive to them, since they extract their coefficients by collecting datasets from multiple measurement campaigns. In addition, both models establish a least-squares problem that involve the $L_2$-norm, highly sensitive to the presence of outliers. Improving the path-loss modelling requires discarding them before trying to compute any model, using robust statistics methods.
3) Atmospheric absorption losses: Atmospheric absorption losses are one of the most important non-linear losses that affect ABG and WABG models, specially oxygen $O_2$ and water $H_2O$. Losses due to both gases non-linearly depend on the carrier frequency, and their loss-curves have been known from decades [46]. A path-loss model should consider gases losses in order to remove their effects and predict accurately the losses.

## III. THE PROPOSED EXTENDED WABG MODEL

To overcome these ABG and WABG models issues, we propose the Extended Weighted ABG model, or simply EWABG, the first non-linear propagation path-loss model which incorporates modern techniques to make it robust against outliers, and to consider the atmospheric gases losses. The proposed model presents important features as in its predecessors: it has a reduced number of coefficients and it applies weighting policies to balance input datasets to be integrated.

Figure 1 shows a flowchart of the main steps to compute the EWABG. Briefly, it is composed of three main steps or blocks: the non-linear EWABG step (red block in Figure 1), which provides a flexible model to overcome the non-homogeneous path-loss media, the robustness block (blue block in Figure 1), which removes possible outliers in raw data, and the atmospheric gases block (yellow blocks in Figure 1) which remove the atmospheric loss effect from the data.

### A. NON-LINEAR EWABG STEP

EWABG model proposes the following second-order non-linear model that expresses the path-losses in terms of 6 parameters ($\alpha_2, \beta, \gamma_2, \alpha_2, \delta, \gamma_2$):

$$P_{\text{EWABG}}(d,f)[dB] = 10\alpha_1 \log_{10}(d) + \beta + 10\gamma_1 \log_{10}(f) \\ + 100\alpha_2 \log_{10}^2(d) + 100\gamma_2 \log_{10}^2(f) \\ + 100\delta \log_{10}(d) \log_{10}(f) + \xi. \quad (6)$$

The $\alpha_1$, $\gamma_1$ and $\beta$ coefficients correspond to the original ABG and WABG models' $\alpha$, $\beta$ and $\gamma$ coefficients (see Equation (1)). Specifically, $\alpha_1$ and $\gamma_1$ are the coefficients that linearly relate the losses to the log-distance and log-frequency, respectively, and $\beta$ is an offset term. The rest of the coefficients $\alpha_2$, $\gamma_2$ and $\delta$ are exclusively introduced by the EWABG, and correspond to the second-order terms of the path-loss power series. Specifically, $\alpha_2$ and $\gamma_2$ relate respectively to the second-order power of the log-distance and log-frequency with the path-losses, and $\delta$ measures the losses' dependence on the cross-product of the log-distance and log-frequency. The term $\xi$, inherited from the ABG model, is a random variable used to represent the path-losses fluctuations, with zero mean and finite variance $\sigma$. Transmitter-receiver distance $d$ is expressed in meters (m) and carrier frequency $f$ in Gigahertz (GHz). Observe that EWABG model is equal to ABG model if the second-order coefficients $\alpha_2$, $\gamma_2$ and $\delta$ are zero.

As it can be seen in Eq. (6), the EWABG model includes all terms up to Taylor series second-order of the generic propagation path-loss function $P(f, d)$, which involves 6 coefficients. As we will see in Section IV, the second-order approximation is enough to absorb the inconsistencies caused by non-homogeneous propagation medium, that cause problems in the ABG and WABG models, especially between high- and low-frequencies, and improves the path-loss estimations in the 5G working range. Higher-order non-linear models have been ruled out since they require to estimate 10 or more coefficients (this number grows polynomially ) and, in Section IV-A, we experimentally show that a third-order non-linear model performs worse than the second-order one (as overfitting appears).

The estimation of the six EWABG parameters is also carried out by establishing a least squares problem, including weighting the residuals according to a weighting policy similar to that in the WABG models, see Section II-B. The best coefficient estimation has the following formula:

$$\begin{pmatrix}\hat{\alpha}_1 \\ \hat{\beta} \\ \hat{\gamma}_1 \\ \hat{\alpha}_2 \\ \hat{\delta} \\ \hat{\gamma}_2\end{pmatrix} = (X^T W X)^{-1} X^T W Y, \quad (7)$$

where $W$ is the same weighting matrix used in WABG (see Section II-B), $Y$ is the dataset path-losses measurements column vector, and $X$ is the distance-frequency data array, built as follows:

$$X = \begin{pmatrix} \mathcal{L}[d_1] & 1 & \mathcal{L}[f_1] & \mathcal{L}^2[d_1] & \mathcal{L}[d_1]\mathcal{L}[f_1] & \mathcal{L}^2[f_1] \\ \mathcal{L}[d_2] & 1 & \mathcal{L}[f_2] & \mathcal{L}^2[d_2] & \mathcal{L}[d_2]\mathcal{L}[f_2] & \mathcal{L}^2[f_2] \\ \vdots & \vdots & \vdots & \vdots & \vdots & \vdots \\ \mathcal{L}[d_i] & 1 & \mathcal{L}[f_i] & \mathcal{L}^2[d_i] & \mathcal{L}[d_i]\mathcal{L}[f_i] & \mathcal{L}^2 f_i] \\ \vdots & \vdots & \vdots & \vdots & \vdots & \vdots \\ \mathcal{L}[d_n] & 1 & \mathcal{L}[f_n] & \mathcal{L}^2[d_n] & \mathcal{L}[d_n]\mathcal{L}[f_n] & \mathcal{L}^2[f_n] \end{pmatrix}, \quad (8)$$

where $\mathcal{L}[\cdot] = 10\log_{10}(\cdot)$. The dimension of $X$ is $n \times 6$.

### B. EWABG ROBUSTNESS STEP

As it has been previously stated, the EWABG model removes the outliers from the raw input measurements (thus eliminating the negative effects on path-loss estimation) by using least-squares closed form formulas, such as the EWABG coefficient estimation in Eq. (8). To carry out this process, we have compared 5 different robust estimators and techniques: Ridge regressor, Lasso regressor, Elasticnet regressor, the RANdom SAmple Consensus (RANSAC) and the Theil-Sen estimator. We briefly describe here these estimators:

- **Ridge [48].** The Ridge regression model minimizes the squared of the residuals as in a simple linear regression, but adds a penalty to the cost function, the $L_2$ norm of the $\boldsymbol{\omega} = (\omega_1, \ldots, \omega_p)$ parameters to be estimated, as follows:

$$\min_{\boldsymbol{\omega}} \sum_{i=1}^{n} (y_i - f(\boldsymbol{\omega}; x_i))^2 + \lambda \|\boldsymbol{\omega}\|_2^2, \quad (9)$$

where $n$ is the number of samples in the dataset. The hyperparameter $\lambda$ is tuned to balance both terms, usually scanning values from 0 to 1. Observe that if $\lambda$ is set to 0, the resulting regressor is simply the linear estimator. Ridge regressor is frequently used to reduce the number of relevant features, removing those that are most correlated among them. Fortunately, as in the linear regression problem, we can find a closed-form expression for the $\omega$ coefficients that minimize Equation (9), which is expressed as follows [48]:

$$\boldsymbol{\omega} = (X^T X + \lambda I)^{-1} X^T Y. \quad (10)$$

In our case, $\boldsymbol{\omega}$ are the EWABG' model coefficients.

- **Lasso [49].** It is the acronym of *Least Absolute Shrinkage and Selection Operator*. As in the Ridge regressor, it establishes a minimization problem that involves the residuals of the model and also an additional penalty in the loss function but substitutes the $L_2$ norm for $L_1$ norm, as follows:

$$\sum_{i=1}^{n} (y_i - f(\boldsymbol{\omega}; x_i))^2 + \lambda \|\boldsymbol{\omega}\|_1 \quad (11)$$

Lasso is used mainly for two purposes: it selects the most relevant regression features and it is more robust against outliers than the Ridge regressor, as the $L_1$ is less sensitive to outliers than the $L_2$. There exists a closed-form regression for but it is frequently to use the Alternating Direction Method of Multipliers algorithm for computing the coefficients [50].

- **ElasticNet [51]** Elasticnet combines both $L_1$ and $L_2$ penalties in the minimization problem as follows:

$$\sum_{i=1}^{n}(y_i - f(x_i))^2 + \lambda_1\lambda_2\|\boldsymbol{\omega}\|_1 + (1-\lambda_1)\lambda_2\|\boldsymbol{\omega}\|_2, \quad (12)$$

where the hyperparameter $\lambda_1$ balances from Ridge to Lasso norms ($L_2$ and $L_1$ norms respectively) varying term from 0 to 1 respectively. Note that if $\lambda_1$ is 0 Elasticnet becomes a Ridge regressor, and if $\lambda_1$ is 1 it becames a Lasso regressor. The hyperparameter $\lambda_2$ is set to balance the penalties terms and the squared of the residuals term.
- **RANSAC [52].** RANdom SAmple Consensus (RANSAC) is an iterative algorithm which directly provides a Machine Learning (ML) model from data in presence of outliers. Briefly, in each iteration, RANSAC chooses a small subset of samples from the dataset and trains an ML-based model with it. Then, the method checks out whether the whole dataset is consistent with the trained model according to a loss function. All samples which fit the trained ML model well (according to the loss function) are part of the so-called consensus set. The other samples that exceed a maximum deviation attributable to the effect of noise are considered outliers. These two steps are repeated a fixed number of times. For each iteration, a ML-model can be rejected if too few points are part of the consensus set, or refined enlarging the corresponding consensus set.
- **Theil-Sen [53].** The Theil-Sen is a a robust regressor which retrieves its coefficients from the median of a set of regressors, directly computed from subsets of samples. Particularly, Theil-Sen computes the median of all the regressions calculated from every pair of points. The algorithm is strongly robust against outliers, and it will conform our robust step in the EWABG. The algorithm is time consuming if the number of samples in the datasets and model parameters is large, which is not our case.

We have used `MATLAB` to implement these regressors, specifically functions: `ridge, lasso, elasticnet, fitPolynomialRANSAC` and `TheilSen`. This last function provided in [53].

### C. THE ATMOSPHERIC GASSES EFFECT

EWABG takes into account the atmosphere gases absorption, which is highly non-linear, although a known effect. Mainly, both $H_2O$ and $O_2$ gases are involved in the propagation path losses and their attenuations depends on the frequency. The International Telecommunication Union (ITU) recommendations, specifically the ITU P.676-12 [46], provide figures (such as Figure 4) that measure the atmospheric attenuation of both types of gases.

Taking advantage of this information, and in order to reduce the amount of non-linear effects, we can subtract the value of the loss produced by the atmosphere's gases at the distance and frequency where the measurements took place (see the first yellow block in Figure 1). Then, after applying the robustness block and estimating the EWABG coefficients, we have to add again the gases loss at this frequency and distance. The value of this absorption is obtained from ITU's recommendation [46].

### IV. COMPUTATIONAL EXPERIMENTS AND RESULTS

In this section, we compare the performance of the EWABG model with that of the state-of-the-art ABG models computed from several 5G datasets. Specifically, from those provided by the companies Nokia and the University of Aalborg (AAU) [45], Qualcomm, the University of Aalto (ALU) and New York University (NYU) [31]; and with state-of-art WABG models provided by [25]. We distinguish between three different types of 5G scenarios: Urban Micro-cellular Street Canyon (UMiSC), Urban Micro-cellular Open Street (UMiOS) and Urban Macro-cellular (UMa). All of them have been considered in a NLOS 5G environment, as in LOS environments the ABG model is accurate enough and presents similar results as the EWABG model.

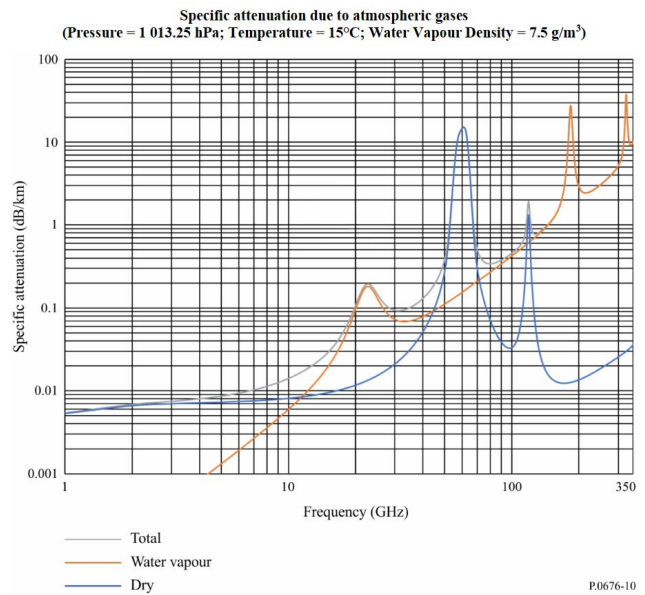

**FIGURE 4.** Specific attenuation due to atmospheric gases. (ITU P.676-12 [46].)

We have carried out three different experiments: The first two experiments determine the best bases combination for the EWABG, i.e., the order of the approximation and the robust estimation method involved, see Sections IV-A and IV-B, respectively. The last experiment is a comparison of the proposed EWABG with the recent state-of-the-art methods, see Section IV-C.

The ABG models for each 5G scenario in the NLOS environment have been obtained from [31]. These ABG models have been individually computed for a single frequency or a stretch frequency band. Table 1 shows the provided ABG coefficients computed from the state-of-the-art datasets in these environments [31]. It includes the number of points used for computing each input ABG, the distance range where the model works and the specific working frequency. Observe that the input models have strong differences in the number of samples they are computed from, and also in their standard deviation. Note that some of them are computed from a low number of points. Although the number of points from which they are calculated are low, they contribute to the overall model, although they should be weighted accordingly. Both WABG and the proposed EWABG weight these models with low number of points, so they can be included in the final outcome of the algorithms.

**TABLE 1.** ABG models available in the literature for UMiSC, UMiOS and UMa 5G scenarios (Sce.) in NLOS environment (Env.) obtained from Nokia/AUU database [45], Qualcomm, Aalto, and NYU databases [31]. The models cover different distance-frequency ranges. Regarding the type: M stands for measurement data and R stands for ray-tracing data. Also, the number of points used to obtain the $\alpha$, $\beta$ and $\gamma$ coefficients is provided, as well as the deviation ($\sigma$) in dB.

| Env. | Sce. | Freq. (GHz) | Company | Nº Points | Dist. Range (m) | Type | $\alpha$ | $\beta$ (dB) | $\gamma$ | $\sigma$ (dB) |
|---|---|---|---|---|---|---|---|---|---|---|
| NLOS | UMiSC | 2 | Nokia/AAU | 27158 | 19-272 | M | 3.5 | 25.0 | 2 | 7.6 |
| | | 2.9 | Qualcomm | 34 | 109-235 | M | 3.9 | 10.2 | 2 | 3.2 |
| | | 18 | Nokia/AAU | 13934 | 19-272 | M | 3.5 | 24.0 | 2 | 8.0 |
| | | 28 | NYU | 20 | 61-186 | M | 2.5 | 51.7 | 2 | 9.7 |
| | | 29 | Qualcomm | 34 | 109-235 | M | 4.2 | 11.0 | 2 | 5.3 |
| | | 73.5 | NYU | 53 | 48-190 | M | 2.9 | 43.2 | 2 | 7.8 |
| | UMiOS | 2 | Nokia/AAU | 10377 | 17-138 | M | 4.7 | −2.2 | 2 | 7.4 |
| | | 2.9 | Qualcomm | 34 | 109-235 | M | 3.9 | 10.2 | 2 | 3.2 |
| | | 18 | Nokia/AAU | 6073 | 23-138 | M | 4.9 | −7.7 | 2 | 7.9 |
| | | 29 | Qualcomm | 34 | 109-235 | M | 4.2 | 11.0 | 2 | 5.3 |
| | | 60 | Aalto | 246 | 8-36 | M | 2.2 | 46.5 | 2 | 1.8 |
| | UMa | 2 | Nokia/AAU | 69542 | 45-1429 | M, R | 3.6 | 7.6 | 2 | 9.4 |
| | | 10.25 | Nokia | 16743 | 45-1174 | R | 2.2 | 47.6 | 2 | 12.5 |
| | | 18 | Nokia/AAU | 27154 | 90-1429 | M | 3.7 | 8.0 | 2 | 6.1 |
| | | 28.5 | Nokia | 16416 | 45-1174 | R | 1.9 | 52.3 | 2 | 12.0 |
| | | 37.625 | NYU | 12 | 61-377 | M | 1.0 | 69.4 | 2 | 9.6 |
| | | 39.3 | Nokia | 16244 | 45-1174 | R | 1.8 | 53.8 | 2 | 11.6 |
| | | 73.5 | Nokia | 15845 | 45-1174 | R | 1.9 | 49.7 | 2 | 10.0 |

Our EWABG model provides a unique path-loss model covering a wide range of frequencies, including all input distance-frequency ranges that the model integrates. Since the number of input ABG models found in [31] is high for the NLOS environment, we will also provide two more EWABG models: one for high frequencies and one for low frequencies, as low and high frequency separation has been established by state-of-the-art methods [21], [22]. Note that this division is not a requirement of the proposed model and we can design a EWABG model in any other frequency range.

As indicated in Table 1, there is a total count of 6, 5 and 7 different ABG models for UMiSC, UMiOS and UMa, respectively. Each model is only valid for the indicated frequency (located between 2 and 73.5 GHz in case of UMiSC and UMa, and between 2 and 60 GHz in case of UMiOS) and covering different distance ranges as indicated.

We use the Weighed Standard Deviation $\sigma$ (WSD) as error metric to evaluate the models, which can be expressed as follows:

$$\sigma = \sqrt{\frac{\sum_{i=1}^{N} W_{ii} (y_i - P_{\text{ABG}}(d_i, f_i))^2}{\sum_{i=1}^{N} W_{ii}}} \quad (13)$$

Note that $\sigma$ is the estimator of the standard deviation of the fluctuation $\xi$ (see Equation (6)). Observe that the WSD directly yields to the simple standard deviation when the weighting matrix is fixed to the identity matrix $W = I$.

### A. DETERMINING THE MODEL's ORDER

We consider two possible order extensions for the proposed EWABG: a second-and a third-order. Higher order models have been discarded, since it is a practical requirement to look for low number of parameters to model the propagation path-loss estimation, and the number of the parameters to be estimated grows polynomially with the model's order (*n*). For second-order models, we have to estimate 6 parameters, see Equation 6. For third-order models, this number increases to 10 parameters and the model has the following expression:

$$\begin{aligned}
P_{EWABG}^{(3)}&(d, f)[dB] \\
&= 10\alpha_1 \log_{10}(d) + \beta + 10\gamma_1 \log_{10}(f) + 10^2 \alpha_2 \log_{10}^2(d) \\
&\quad + 10^2 \delta \log_{10}(d) \log_{10}(f) + 10^2 \gamma_2 \log_{10}^2(f) \\
&\quad + 10^3 \alpha_3 \log_{10}^3(d) + 10^3 \zeta \log_{10}(d) \log_{10}^2(f) \\
&\quad + 10^3 \eta \log_{10}^2(d) \log_{10}(f) + 10^3 \gamma_3 \log_{10}^3(f) + \xi \quad (14)
\end{aligned}$$

We use the UMiSC ABG models provided in Table 1 to construct two second-and third-order EWABG models that integrate them. We determine the best model evaluating both the qualitative behaviour in those areas with no data and the error modelling path-losses in well-known distance-frequency ranges. We do not include outliers in the data, and also we do not remove the gases effect. We use the Nokia/AAU at 18 GHz as a test model (ground truth) in order to evaluate the performance of the EWABG, therefore this model is not integrated. The 18 GHz frequency is located close to the middle of the 5G frequency range, and this model is computed from a large amount of samples, see Table 1. However, we have also developed a leave-one-out cross-validation (LOOCV) with all the UMiSC ABGs for evaluating the performance of the second-and third-order EWABG.

Figure 5 shows the resulting EWABG for both second-and third-order estimations. Figure 6 also shows the path-losses estimated for both second-and third-order EWABG and WABG versus distance and frequency, Figures 6(a) and 6(b), respectively. Figure 6(a) sets the frequency to 18 GHz and

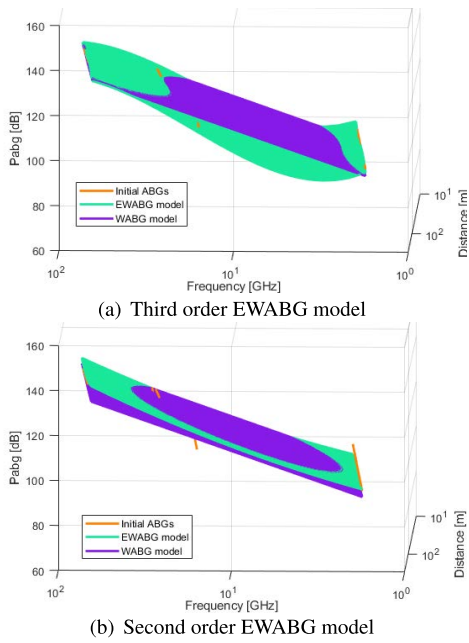

(a) Third order EWABG model

(b) Second order EWABG model

**FIGURE 5.** EWABG's models comparative with different order. (scenario UMiSC 2-73,5 GHz ).

**TABLE 2.** Model deviation $\sigma$ [dB] at 18 GHz and the final results of LOOCV process.

| Model | 3-order EWABG | 2-order EWABG | WABG |
|---|---|---|---|
| 18 GHz | 8.313 | 8.561 | 8.854 |
| LOOCV | 7.327 | 7.772 | 8.107 |

Figure 6(b) sets the distance to 100 m. Table 2 shows the obtained error at a fixed frequency of 18 GHz (ground truth). We also add the resulting WABG model in order to be used as baseline. The estimated third-order EWABG has obtained a standard deviation of 8.31, a slightly lower value than that obtained by the second-order EWABG 8.56 at 18 GHz, due to the increase of parameters. The results of LOOCV have the same trend. The estimated third-order EWABG has obtained a result of 7.33, a slightly lower value than that obtained by the second-order EWABG 7.77 in the LOOCV. However, the qualitative path-losses third-order EWABG estimation shown in Figure 5, presents a strange behaviour, an asymmetric valley, in the frequencies between 1 GHz and 18 GHz. The third-order EWABG is overfitted since the ripples and valleys that appear are not consistent with the physical behavior of the path-losses that always grows with distance and frequency. On the contrary, the second-order EWABG estimates increasing path-losses with distance and frequency, see Figure 6. Therefore, we will use the second-order approximation for the construction of the EWABG model.

### B. ROBUSTNESS ESTIMATION

In order to determine which of the regressors and estimators described in Section III-B (Ridge, Lasso, Elasticnet, RANSAC and Theil-Shen) presents best results in presence of outliers, we use the UMiSC Qualcomm ABG model to generate observations at 2.9 GHz and then contaminate the data with outliers following a Raygleigh distribution (Eq. (15)).

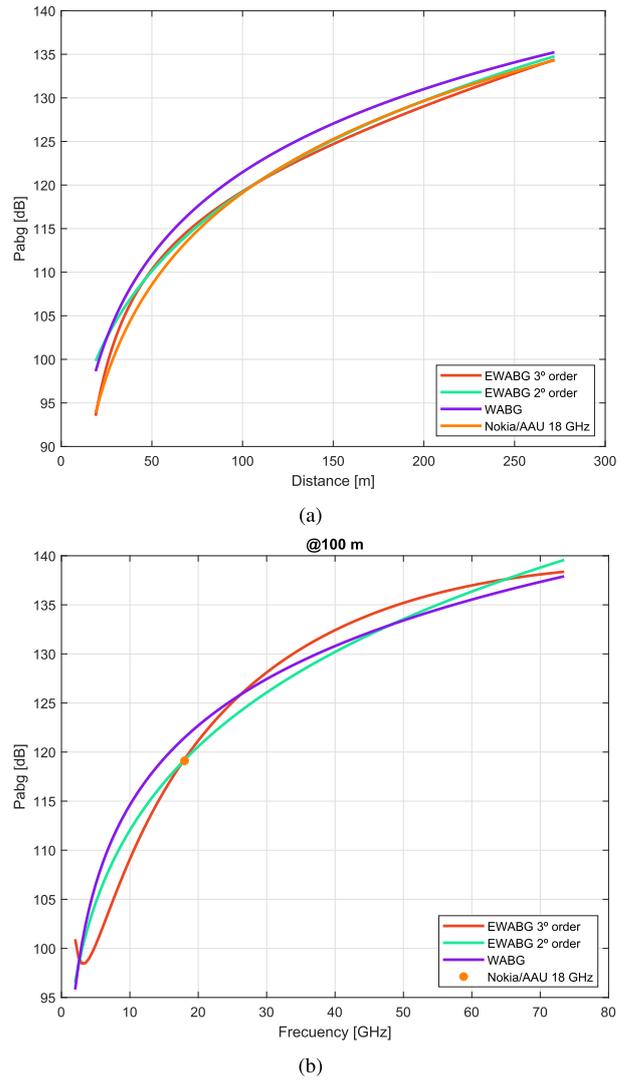

(a)

(b)

**FIGURE 6.** EWABG's models comparative with different order. a) path-losses vs. distance at 18 GHz and b) path-losses vs. frequency at 100 m.

$$f(x) = \frac{x}{\rho} e^{\frac{-x^2}{2\rho}}, \quad x > 0, \quad (15)$$

where $\rho$ is set to 0.75. The Raygleigh distribution is commonly used for modelling scattering noise, frequent in radio wave propagation. We run 10 trials and report the average of the standard deviation model with and without outliers.

Regarding the implementation, in case of Ridge, Lasso and Elasticnet regresors, we use a K-fold cross validation with $K = 10$ to optimize the hyperparameter $\lambda$ in the case of Ridge and Lasso, and $\lambda_1$ and $\lambda_2$ in case of Elasticnet. For RANSAC, we set the number of iterations to 1000 in order to limit the number of operations. Table 3 shows the results obtained from all the robust estimators.

**TABLE 3.** Standard deviation of the robust regressors for the Qualcomm 2.9 GHz UMiSC scenario. The best performances have been highlighted in boldface font.

| Robust Regressor | $\sigma$ with outl. [dB] | $\sigma$ without outl. [dB] |
|---|---|---|
| Lineal | 4.749 | 3.633 |
| Lasso | 4.739 | 3.63 |
| Ridge | 4.69 | **3.623** |
| ElasticNet | 4.711 | 3.624 |
| RANSAC | 4.654 | 3.631 |
| Theil-Sen | **3.902** | 3.634 |

It is possible to see that, without introducing outliers in the observations, all the estimators report similar errors. The best results are reported by the Ridge regresor, but it is closely followed by all the evaluated methods. When outliers are introduced in the experimentation, the best performance has been obtained by the Theil-Sen estimator with 3.902 dB, close to the results obtained without outliers. The other robust estimators reported worse results. RANSAC is placed in the second position reporting 4.654 dB. Theil-Sen directly operates over the median of all possible regressors. Note that statistical estimators such as Theil-Sen are known to be suitable approaches for removing outliers. As a consequence of the results obtained in this Section, we use the Theil-Sen estimator to build our robust regressor, which will remove the possible outliers from the input data.

### C. EXPERIMENTS IN NLOS ENVIRONMENT

NLOS is an environment where phenomena like scattering and fading are more pronounced than in LOS environments, increasing the obtained standard deviation figures. In this section, we provide EWABG models from the existing ABGs, see Table 1, for the UMiSC, UMiOS and UMa scenarios in NLOS environment. We have 6, 5 and 7 different ABG models for UMiSC, UMiOS and UMa. The initial ABG models (see Table 1) are only valid for the indicated frequency (located between 2 and 73.5 GHz in case of UMiSC and UMa, and between 2 and 60 GHz in case of UMiOS) and covering different distance ranges as indicated.

Since we have a considerable large number of NLOS ABG models for the different scenarios (see Table 1), we develop three different models for each case: a low frequency model between 2 and 18 GHz, a high frequency model between 28 and 73.5 GHz and a global model covering the entire mmWave frequency range. In the case of UMiOS, the upper-bound frequency of the generated model has been set to 60 GHz since there is no existing approach that estimates propagation path-losses above 60 GHz in this scenario (see Table 1). We follow the same separation that appears in Sun [31], in order to be able to compare our results with that work.

We carry out a comparison of the proposed EWABG model in two different experiments. The first one does not include outliers in the data. The second one includes outliers in the experimentation to evaluate the global improvement of our EWABG. The comparison includes our EWABG approach together with those by WABG [25], Sun [31] and 3GPP [9]. In order to compare to the 3GPP path-loss model [9], Tables provided in Recommendation ETSI TR 138 901 V15.0.0 have been used. Note that the only scenarios considered in this recommendation are UMiSC and UMa)

Table 4 shows the results obtained by the proposed EWABG and the other state-of-the-art approaches included in the comparison, without adding outliers in the experimentation. We show both the reported standard deviation in the original paper $\sigma_{orig}$ and the one obtained in our experiments under the same conditions $\sigma_{exp}$. We discuss the result for each 5G scenario.

For the UMiSC scenario, the EWABG obtains the best standard deviation results in both low and high frequency range models, and even for the global frequency model with 4.8, 7 and 5.93 dB respectively. It significantly surpasses the WABG and Sun16 models for low frequencies. It clearly improves Sun16's estimation for all frequency ranges, since this model prioritizes model estimations with a larger number of samples. For instance, in the case of the frequency range between 2 GHz and 18 GHz, the EWABG model prioritizes the Nokia/AAU model against the others, see Tables 1 and 4, since the number of samples within this model is higher than in the other databases. The existence of original unbalanced models leads to unbalanced path-loss estimations that are well weighted by the WABG and EWABG. Both EWABG and WABG models contribute to balance the involved original ABG, considering both the standard deviation and the number of observations. Our WABG model reduces the standard deviation in the low, high and global path-loss models.

EWABG model also improves the error obtained by Sun16 for the UMiOS scenario for all low, high and global frequency ranges. The model achieves 4.75 dB and 3.45 dB of standard deviation for low and global frequency ranges, respectively, clearly improving the model obtained by Sun16 in more than 3 dB. In the high frequency ranges, EWABG obtains the best result, but the improvement is smaller. In this case the WABG also obtains good results, close to those obtained by the proposed EWABG.

In the UMa scenario, the EWABG model obtains the best standard deviation, better than WABG and Sun16 for all frequency ranges, except for the global frequency range, in which it is surpassed by 3GPP, with 8.2 dB versus 9.63 dB of the EWABG. However, EWABG, WABG and Sun16 have larger working ranges, covering longer distances than the 3GPP model. We observe that in UMa scenario, differences are not as severe as in UMiSC and UMiOS scenarios.

The experiments carried out reveal a considerable improvement of the proposed EWABG compared to the Sun16 model, above 3 dB in almost all scenarios and frequency ranges. This improvement is mainly caused by the treatment of unbalanced data applied in the proposed model. The EWABG also improves the results obtained by the WABG (which also applies a weighting data step), due to the better estimations reached using a second-order polynomial, and the inclusion of the atmospheric gasses effect in the model. All in all, an improvement of 0.5 dB in the standard deviation is achieved. However, note that the best performance of the proposed EWABG is achieved in presence of outliers, where the rest of path-propagation models fail to estimate path-losses. We will show this effect in the next experiments.

**TABLE 4.** EWABG models for low frequencies, high frequencies and total frequency range in a NLOS environment (Env.) for the different scenarios (Sce.) considered. Comparison is performed against the ABG models by Sun16 [31] and WABG [25].

| Env. | Sce. | Freq. Range (GHz) | Method | Nº Points | Dist. Range (m) | $\sigma_{orig}$ (dB) | $\sigma_{exp}$ (dB) |
|---|---|---|---|---|---|---|---|
| NLOS | UMiSC | 2-18 | Sun16 | 54350 | 19-272 | 8 | 7.95 |
| | | | WABG | 635 | 19-272 | 5.9 | 5.27 |
| | | | EWABG | 635 | 19-272 | - | **4.80** |
| | | 28-73.5 | Sun16 | 107 | 48-235 | 7.8 | 7.88 |
| | | | WABG | 396 | 48-235 | 7.3 | 7.16 |
| | | | EWABG | 396 | 48-235 | - | **7** |
| | | 2-73.5 | 3GPP16 | - | 10-200 | 8.2 | 8.2 |
| | | | Sun16 | 54457 | 19-272 | 8 | 7.9 |
| | | | WABG | 1031 | 19-272 | 6.5 | 6.02 |
| | | | EWABG | 1031 | 19-272 | - | **5.93** |
| | UMiOS | 2-18 | Sun16 | 21888 | 17-235 | 7.6 | 7.47 |
| | | | WABG | 365 | 17-235 | 5.2 | 4.91 |
| | | | EWABG | 365 | 17-235 | - | **4.75** |
| | | 29-60 | Sun16 | 280 | 8-235 | 2.6 | 2.62 |
| | | | WABG | 156 | 8-235 | 2.5 | 2.5 |
| | | | EWABG | 156 | 8-235 | - | **2.41** |
| | | 2-60 | Sun16 | 22168 | 8-235 | 7.8 | 7.83 |
| | | | WABG | 521 | 8-235 | 3.8 | 3.9 |
| | | | EWABG | 521 | 8-235 | - | **3.45** |
| | UMa | 2-18 | Sun16 | 137981 | 45-1429 | 8.9 | 8.84 |
| | | | WABG | 3855 | 45-1429 | 8.4 | 8.36 |
| | | | EWABG | 3855 | 45-1429 | - | **8.35** |
| | | 28.5-73.5 | Sun16 | 48517 | 45-1174 | 11.2 | 11.18 |
| | | | WABG | 3707 | 45-1174 | 10.9 | 10.73 |
| | | | EWABG | 3707 | 45-1174 | - | **10.68** |
| | | 2-73.5 | 3GPP16 | - | 10-1000 | **8.2** | **8.2** |
| | | | Sun16 | 186498 | 45-1429 | 9.9 | 9.86 |
| | | | WABG | 7562 | 45-1429 | 9.6 | 10.09 |
| | | | EWABG | 7562 | 45-1429 | - | 9.63 |

For this experiment we use the same outliers' distribution as in Section IV-B, considering three different cases depending on the outliers dispersion: 50 m, 30 m and 5 m. All theses cases simulate different levels of scattering noise. The multi-path noise is mainly caused by the phase and amplitude distortion generated when the frequency waves reflect off different sized buildings. Therefore, the outliers have been placed around the center of the distance range, simulating an obstacle. Table 5 shows the results obtained by the proposed EWABG and the other state-of-the-art models presented, when adding outliers to the experimentation. Note that this table provides the obtained standard deviation values in the three studied cases $\rho 50m$, $\rho 30m$ and $\rho 5m$, respectively, and the incremental error ratio compared to the cases without outliers. As the deviation increases, the estimation will be worse, hence, good models should have incremental error rates close to 0.

In general, the EWABG obtains an error ratio close to 0 in almost all evaluated 5G scenarios, for all working ranges and all outliers cases. In case of having outliers in a band of 50 m, the EWABG obtains an increment error ratio lower than 1% in almost all cases, compared to the original case, except for the UMiSC in high frequencies where it obtains a 1.85%. Note that the EWABG method is able to remove the outliers effect, obtaining similar standard deviations in presence and absence of outliers. Indeed, the negative figures reported in the EWABG column reflect improvements as some observations have been removed by the model. Also note that the EWABG does not report better results than in the case of no presence of outliers, which is consistent with the existence of an environmental and modeling noise, intrinsic to the problem. Both WABG and Sun16 report worse results, since they are not able to remove the outliers' effect. Specifically, Sun16 strongly suffers from the outlier effects, reporting error ratios above 10%, specially in UMiSC and UMiOS. WABG in less sensitive that Sun16 to outliers, but it still presents poor performance with error rates above 5% in UMiSC and UMiOS. In UMa scenario, all methods report good performance against outliers. The reason is that in UMa scenario, distance ranges are larger than in UMiSC or UMiOS, and hence, the outliers effect diminishes in this case. When we consider an outliers band of 30 m (equivalent to consider a smaller obstacle), the EWABG again obtains good results, with an increment error ratio lower than 1% in almost all cases, except for the high frequencies UMiSC, where the model obtains 4.86%. Therefore, modifying the obstacle size does not affect the EWABG performance, which is able to remove the outliers' effect. Again, Sun16 and WABG report much larger increments of error rates. Sun16 obtains error rates above 10%, similar values compared to the the previous outlier case in UMiSC, but worse error in UMiOS, specially in high frequency ranges. WABG also obtains worse results compared to the previous outliers case, but inverting this behavior: it improves in UMiSC but it is worse in UMiOS scenarios. Finally, we experiment by adding outliers in a fine band of 5 m. This case introduces less outliers, but focused in a very narrow area. The EWABG obtains increment error ratios close to 0, the best results compared to Sun16 and WABG. However, we observe that both Sun16 and EWABG obtain good results too around 1% and 2%, respectively. In this case, the outliers are not so frequent to significantly change the standard deviation compared to the case without outliers, which explains the good results obtained by Sun16 and EWABG.

**TABLE 5.** EWABG models for low frequencies, high frequencies and total frequency range in a NLOS environment (Env.) for the different scenarios (Sce.) considered in presence of outliers (rayleigh distribution with $\sigma_{\cdot m}$). Comparison is performed against the ABG model by Sun16 [31] and WABG [25].

| Env. | Sce. | Freq. Range (GHz) | Method | $\sigma_{50m}$ (dB) | Err. Rat. (%) | $\sigma_{30m}$ (dB) | Err. Rat. (%) | $\sigma_{5m}$ (dB) | Err. Rat. (%) |
|---|---|---|---|---|---|---|---|---|---|
| NLOS | UMiSC | 2-18 | Sun16 | 8.94 | 13.39 | 8.83 | 12.11 | 7.94 | 0.74 |
| | | | WABG | 5.75 | 7.81 | 5.7 | 7.04 | 5.58 | -1.84 |
| | | | EWABG | **5.09** | **−0.73** | **5.09** | **−0.6** | **4.91** | **−1.97** |
| | | 28-73.5 | Sun16 | 8.89 | 13.79 | 8.95 | 14.71 | 7.96 | 2.03 |
| | | | WABG | 7.76 | 11.39 | 7.92 | 13.79 | 7.05 | 1.30 |
| | | | EWABG | **7.07** | **1.85** | **7.28** | **4.86** | **7.01** | **0.99** |
| | | 2-73.5 | Sun16 | 9.08 | 13.49 | 9 | 12.52 | 8 | 0 |
| | | | WABG | 6.29 | 5.94 | 6.38 | 7.46 | 6.06 | 2.01 |
| | | | EWABG | **5.91** | **0.9** | **5.97** | **1.96** | **5.97** | **1.97** |
| | UMiOS | 2-18 | Sun16 | 8.58 | 16.4 | 8.73 | 18.39 | 7.57 | 2.65 |
| | | | WABG | 5.58 | 12.59 | 5.47 | 10.28 | 4.93 | -0.44 |
| | | | EWABG | **4.89** | **−0.11** | **4.88** | **−0.39** | **4.86** | **−0.82** |
| | | 29-60 | Sun16 | 3.73 | 47.57 | 3.92 | 54.79 | 2.59 | 2.3 |
| | | | WABG | 3.09 | 22.49 | 2.99 | 18.47 | 2.87 | 13.93 |
| | | | EWABG | **2.47** | **0.71** | **2.43** | **−0.87** | **2.39** | **−2.6** |
| | | 2-60 | Sun16 | 8.73 | 12.21 | 8.84 | 13.68 | 8.48 | 1.25 |
| | | | WABG | 4.24 | 10.35 | 4.17 | 8.67 | 3.84 | 7 |
| | | | EWABG | **3.45** | **1** | **3.45** | **1** | **3.51** | **2.83** |
| | UMa | 2-18 | Sun16 | 9 | 1.82 | 8.88 | 0.49 | 8.67 | 0.93 |
| | | | WABG | 8.38 | -0.48 | 8.35 | -0.87 | 8.35 | -0.88 |
| | | | EWABG | **8.33** | **−0.87** | **8.32** | **−0.9** | **8.33** | **−0.82** |
| | | 28.5-73.5 | Sun16 | 11.37 | 2.03 | 11.25 | 0.98 | 11.12 | -0.2 |
| | | | WABG | 11.04 | 2.67 | 10.87 | 1.15 | 10.74 | 0 |
| | | | EWABG | **10.79** | **0.24** | **10.75** | **−0.12** | **10.7** | **-0.6** |
| | | 2-73.5 | Sun16 | 9.97 | 1.08 | **9.88** | **0.24** | 9.91 | 0.55 |
| | | | WABG | 10.18 | 0.56 | 10.19 | 0.72 | 10.11 | -0.11 |
| | | | EWABG | **9.90** | **−0.16** | 9.96 | 0.48 | **9.91** | **0.01** |

**TABLE 6.** Coefficients obtained for the EWABG, WABG [25] and Sun16 [31] models.

| Env. | Sce. | Freq. Range (GHz) | Method | $\alpha_1/\alpha$ | $\beta$ | $\gamma_1/\gamma$ | $\alpha_2$ | $\delta$ | $\gamma_2$ | $\sigma$ (dB) |
|---|---|---|---|---|---|---|---|---|---|---|
| NLOS | UMiSC | 2-18 | Sun16 | 3.5 | 24.45 | 1.9 | - | - | - | 7.95 |
| | | | WABG | 3.17 | 26.21 | 2.3 | - | - | - | 5.27 |
| | | | EWABG | 2.35 | 51.7 | −4.8 | 0.03 | −0.01 | 0.45 | **4.80** |
| | | 28-73.5 | Sun16 | 2.7 | 35.95 | 2.60 | - | - | - | 7.88 |
| | | | WABG | 2.8 | 35.2 | 2.54 | - | - | - | 7.16 |
| | | | EWABG | −8.66 | 1985.1 | −221.82 | 0.27 | 0.05 | 6.72 | **7** |
| | | 2-73.5 | Sun16 | 3.5 | 24.48 | 1.9 | - | - | - | 7.92 |
| | | | WABG | 3.16 | 24.61 | 2.68 | - | - | - | 6.02 |
| | | | EWABG | −0.07 | 62.06 | 1.04 | 0.08 | 0.02 | 0.06 | **5.93** |
| | UMiOS | 2-18 | Sun16 | 4.97 | −2.89 | 1.77 | - | - | - | 7.47 |
| | | | WABG | 3.92 | 10.54 | 1.96 | - | - | - | 4.92 |
| | | | EWABG | 7.11 | −9.46 | −3.47 | −0.07 | 0.03 | 0.3 | **4.75** |
| | | 29-60 | Sun16 | 2.4 | 74.06 | 0.31 | - | - | - | 2.62 |
| | | | WABG | 2.62 | 93.75 | −0.7 | - | - | - | 2.5 |
| | | | EWABG | 16.35 | −277.37 | 25.56 | −0.02 | −0.77 | −0.31 | **2.41** |
| | | 2-60 | Sun16 | 4.39 | 2.72 | 1.89 | - | - | - | 7.83 |
| | | | WABG | 3.04 | 26.93 | 2.49 | - | - | - | 3.9 |
| | | | EWABG | −1.97 | 71.39 | −0.24 | 0.15 | 0.03 | 0.11 | **3.45** |
| | UMa | 2-18 | Sun16 | 3.6 | 7.56 | 2.4 | - | - | - | 8.84 |
| | | | WABG | 3.43 | 11.17 | 2.34 | - | - | - | 8.36 |
| | | | EWABG | 1.49 | 35.54 | 2.39 | 0.04 | −0.01 | 0.01 | **8.35** |
| | | 28.5-73.5 | Sun16 | 1.9 | 64.6 | 1.21 | - | - | - | 11.18 |
| | | | WABG | 1.96 | 55.01 | 1.62 | - | - | - | 10.73 |
| | | | EWABG | −0.79 | 199.34 | −11.46 | 0.06 | −0.01 | 0.39 | **10.68** |
| | | 2-73.5 | Sun16 | 3.3 | 17.53 | 2 | - | - | - | 9.86 |
| | | | WABG | 2.86 | 31.58 | 1.74 | - | - | - | 10.09 |
| | | | EWABG | −0.67 | 49.61 | 6.49 | 0.09 | −0.1 | −0.1 | **9.63** |

## V. CONCLUSION
In this paper we have proposed a robust non-linear generalized path-loss propagation model, called *EWABG*. The model is able to integrate different available datasets acquired in multiple distance-frequency ranges and other 5G propagation path-losses models from the literature. The EWABG is the first non-linear extension of the ABG-based approach, that allows overcoming the non-uniformity problem between the high and low 5G frequencies. In addition, the EWABG also addresses the problem of removing the outliers effect, which has not been included before in previous path-loss models. We have compared the most recent techniques for avoiding outliers, and we have adopted the Theil-Sen method to be incorporated in the EWABG, due to its strong robustness demonstrated in the experiments carried out. In addition, the EWABG specifically considers non-linear attenuation by atmospheric gases, in order to improve its estimations. The good performance of the proposed EWABG model has been evaluated in different experiments, and we have carried out a direct comparison with recently proposed 5G propagation path-loss models, including the ABG and WABG, which include 5G non-line-of-sight environment in several 5G scenarios, UMiSC, UMiOS and UMa. EWABG reports 5.93, 3.45 and 9.63 dBs of standard deviation in UMiSC, UMiOS and UMa, respectively, outperforming both ABG and WABG models. Also, the standard deviation error rate reported in presence of outliers is around 0.1% to 2%, which proves its strong outliers rejection capacity. We finally have reported the parameters of the EWABG for its possible use in practical applications.

## APPENDIX. COEFFICIENTS OF THE MODELS OBTAINED
Table 6 shows the coefficients of the proposed EWABG as well as WABG [25] and Sun16 [31] models, integrating the ABGs of Table 1.

## ACKNOWLEDGMENT
The authors would like to thank Juan Antonio Carabaña, José Luis Ruiz-Mendoza, and Sonia Castillo for their support and assessment. They would also like to thank the Secretaría de Estado de Telecomunicaciones e Infraestructuras Digitales, specifically D. Antonio Fernández-Paniagua, for their guidance and problem definition.

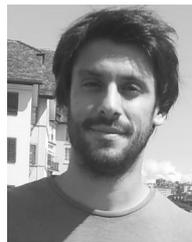

**DAVID CASILLAS-PÉREZ** received the B.S. degree in telecommunication engineering, in 2013, and the M.S. and Ph.D. degrees in electronic control systems from Universidad de Alcalá, Spain, in 2013, 2014, and 2019, respectively. He is currently an Assistant Professor with the Department of Signal Processing and Communications, Universidad Rey Juan Carlos, Madrid, Spain. His research interests include the development of machine learning algorithms with applications in different fields, such as computer vision, mobile communication systems, and renewable energy systems.

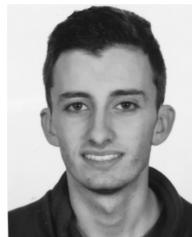

**DANIEL MERINO-PÉREZ** was born in Madrid, Spain, in 1997. He received the degree in telecommunication technologies engineering and the master's degree in telecommunication engineering from Universidad de Alcalá, in 2019 and 2021, respectively. His research interests include mobile communication systems and SoC's architectures.

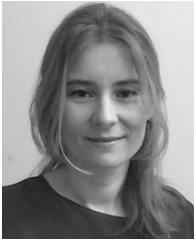

**SILVIA JIMÉNEZ-FERNÁNDEZ** was born in Madrid, Spain, in 1976. She received the B.S. and Ph.D. degrees in telecommunications engineering from Universidad Politécnica de Madrid, Spain, in 1999 and 2009, respectively. She is currently an Associate Professor with the Department of Signal Processing and Communications, where she carries out research on the application of signal processing and machine learning techniques for mobile communication systems.

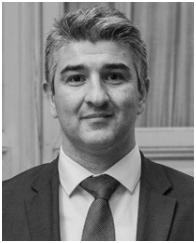

**JOSE A. PORTILLA-FIGUERAS** was born in Santander, Spain, in 1976. He received the B.S. and Ph.D. degrees in telecommunications engineering from Universidad de Cantabria, Spain, in 1999 and 2004, respectively. He is currently a Full Professor with the Department of Signal Processing and Communications, and the Head of the Polytechnic School of Universidad de Alcalá, Spain. His current research interests include mobile communications systems, 5G systems, and the development of machine learning algorithms with application in telecommunication engineering problems.

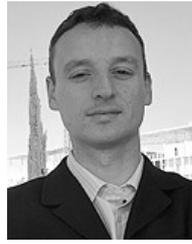

**SANCHO SALCEDO-SANZ** was born in Madrid, Spain, in 1974. He received the B.S. degree in physics from Universidad Complutense de Madrid, in 1998, the Ph.D. degree in telecommunications engineering from the Universidad Carlos III de Madrid, in 2002, and the Ph.D. degree in physics from Universidad Complutense de Madrid, in 2019. He was a Postdoctoral Research Fellow at the School of Computer Science, The University of Birmingham, U.K. Currently, he is a Full Professor with the Department of Signal Processing and Communications, Universidad de Alcalá, Spain. He has coauthored more than 225 international journal articles in the field of machine learning and soft-computing. His current research interests include soft-computing techniques, hybrid algorithms, and neural networks in different applications of science and technology.

○ ○ ○